\documentclass[showpacs,reprint,groupedaddress]{revtex4-1}
\usepackage{color}
\usepackage{setspace}
\usepackage{verbatim}
\usepackage{graphicx}
\usepackage{subfigure}
\usepackage{amsmath}
\usepackage{soul}
\usepackage{multirow}
\usepackage{bbm} 
\definecolor{orange}{rgb}{1.0,0.76,0.02}

\setstcolor{blue}

\begin{document}

\title{Maximum Likelihood Analysis of Reaction Coordinates during Solidification in Ni}

\author{Grisell D\'{i}az Leines}
\email{grisell.diazleines@rub.de}
\affiliation{Interdisciplinary Centre for Advanced Materials Simulation, Ruhr-Universit{\"a}t Bochum, 44780 Bochum, Germany}
\author{Jutta Rogal}
\affiliation{Interdisciplinary Centre for Advanced Materials Simulation, Ruhr-Universit{\"a}t Bochum, 44780 Bochum, Germany}

\begin{abstract}
Understanding the underlying mechanism of crystal nucleation 
during solidification is a fundamental aspect in the prediction and control of
materials properties. Classical nucleation theory (CNT) assumes that homogeneous nucleation occurs
via random fluctuations within the supercooled liquid, that the structure of the growing clusters resembles the most stable 
bulk phase, and that the nucleus size is the sole reaction coordinate (RC) of the process. 
Many materials are, however, known to exhibit multiple steps during crystallization, forming different polymorphs. 
As a consequence,  more complex RCs are often required to capture all relevant information about the process.
In this work, we employ transition path sampling together with a maximum likelihood analysis of candidate order 
parameters to identify suitable reaction coordinates for the nucleation mechanism during solidification in Ni. In contrast to CNT, the analysis
of the reweighted path ensemble shows that a pre-structured liquid region that surrounds the crystal cluster is a relevant order parameter that
enhances the RC and therefore plays a key role in the description of the growing nucleus and the interfacial 
free energy.  We demonstrate that pre-structured liquid clusters that emerge within the liquid act as precursors of the crystallization
in a non-classical two-step mechanism which predetermines the coordination of the polymorphs that are being selected.

\end{abstract}
\maketitle


\section{Introduction}

Crystal nucleation is a fundamental 
process of solidification observed 
in a large variety of phenomena across 
different disciplines from physics to biology. 
Despite its importance,  understanding the 
mechanism of crystal nucleation on the 
atomistic level remains elusive due to 
the time and length scales of the process that hamper 
experimental and computational techniques ~\cite{Sosso2016, Anwar2011, Jungblut2016}.
For atomistic simulations, the identification of a meaningful 
reaction coordinate (RC) is a crucial step to 
understand the mechanism of nucleation, but 
the high dimensionality and extended timescales 
of the process often pose a major challenge 
in the identification of order parameters that define the RC. 

Classical nucleation theory (CNT)~\cite{Becker1935,Binder1987}
provides a simple, phenomenological description of the 
nucleation mechanism, where the free energy of 
a growing crystalline nucleus is described by the 
competition between 
a volume and a surface term.
The volume term leads to a decrease in energy associated with the difference in chemical potential between the solid and the liquid phase, whereas the surface term increases the energy due to the formation of a solid-liquid interface.
An essential assumption of CNT is that nucleation is 
a one-dimensional process, i.e. described 
by a single RC, namely the radius of 
the spherical cluster or the size of the crystal nucleus.  
Other core assumptions  of CNT are
that small clusters of  hundreds of particles 
emerge randomly within the liquid, 
having the same thermodynamic properties (the capillarity approximation) and crystal structure as the bulk.
However, atomistic simulations that allow to follow different 
order parameters along the nucleation pathway or sample 
multi-dimensional free energy landscapes 
have revealed a variety of examples where nucleation can proceed via the 
ordering of more than one order parameter,~\cite{Moroni2005, Trudu2006, Lechner2011, Peters2006,Tanaka2016} 
and where the polymorphs  that are formed do not always correspond to the thermodynamically most stable phase of the bulk.~\cite{Lechner2011, PhysRevE.76.031604, Beckham2011, Jungblut2013, tenWolde1997}
Recently, two-step non-classical nucleation mechanisms have raised great interest 
and were found in several systems including  hard spheres, 
colloidal systems, globular proteins and metallic systems.~\cite{tenWolde1997, tenWolde1999, PhysRevLett.105.025701,Kawasaki2011,Russo2012,Russo2012b, Tanaka2016,DLeines2017}
In these cases, the nucleation occurs via the formation of 
regions within the undercooled liquid of either high density or high orientational order that 
act as seeds or precursors of the crystallization. A pre-ordered surface region  
that embeds a crystal core was recently shown to enhance the RC for nucleation in colloidal suspensions~\cite{Lechner2011a} resulting in a two-dimensional description 
of nucleation process and a reinterpretation of the surface-volume variables within CNT.

Our previous findings indicate that homogeneous nucleation in Ni also occurs via a 
two-step crystallization process where long-lived
mesocrystal regions of high 
orientational order mediate the nucleation of the crystal phase that grows embedded 
in a pre-structured liquid cloud.~\cite{DLeines2017} 
But it remains still unclear 
if the pre-structured region triggers the initial nucleation in the  Ni melt
and which role it plays in defining the RC of the process. In the present study 
we approach these open questions and investigate 
the formation of different crystal structures
with a multi-dimensional analysis of the microscopic pathways for nucleation in Ni. We employ 
statistical path ensembles from transition path sampling (TPS) simulations~\cite{Dellago2002, VanErp2005} 
and perform a quantitative 
analysis of the RC  and the free energy landscape of the nucleation process  
by maximum likelihood estimation (MLE).~\cite{Peters2006, Rogal2010, Lechner2010} 
The quality of different order parameters as RC is evaluated by their ability to model the committor data obtained from the path ensemble, where the committor or commitment probability is a statistical measure of the progress of a reaction.
In particular, we employ the MLE approach to extract a non-linear description of the RC from the reweighted path ensemble (RPE).~\cite{Lechner2010,Rogal2010} 
We show that the pre-structured liquid cloud significantly enhances the description of the RC,  
whereas the size of the cluster composed only of face-centered cubic (fcc) atoms, i.e. atoms with the structure of the final bulk phase as suggested by CNT,  provides a much less suitable RC.
Including the pre-structured 
cloud in the RC furthermore improves the description of the interfacial free energy between the growing nucleus and the liquid, 
resolving a discrepancy between the nucleation barrier (and associated rate constant) 
obtained by CNT and experiments.~\cite{Bokeloh2011}  
The unbiased free energy landscape obtained from the RPE~\cite{Bolhuis2011,Rogal2010} reveals  a non-classical nucleation pathway for solidification in Ni.  
The nucleation proceeds via the initial formation of a long-lived mesocrystal region 
of higher orientational order than the liquid but less symmetry than the crystal structures.
Subsequently, the crystalline phase emerges within the core of this region and prevails in the core of the growing cluster.
The mesocrystal region is mainly composed of distorted polyhedra with 12 vertices that resemble the coordination polyhedra in 
fcc and hexagonal close-packed (hcp) structures, indicating that the pre-ordered regions within the liquid 
pre-determine the polymorph selected during the crystallization, therefore acting as a precursor 
of the nucleation process. 

This paper is organized as follows: in Sec.~\ref{methods} we present the methods used to analyze the RC of
nucleation during solidification in Ni; in Sec.~\ref{tis_simulations} we discuss the nucleation mechanism obtained
from TPS simulations and possible candidate order parameters for the RC; in Sec.~\ref{RCanalysis} we present the MLE analysis of
the RC and we discuss the role  of the pre-structured liquid cloud in the nucleation mechanism; we conclude our findings
in Sec.~\ref{conclusion}.


\section{Methods}
\label{methods}


\subsection{Transition interface sampling}

Crystal nucleation is an activated process where a large free energy barrier separates the melt from the solid  state. 
To sample the transition from liquid to solid during nucleation in Ni requires an advanced rare event method. Here, we employ transition interface sampling (TIS),~\cite{VanErp2005} a method that allows to  explore an ensemble of phase space trajectories (or paths) between two stable states $A$ and $B$ (liquid and solid) using a Monte Carlo (MC) framework in trajectory space. While the original transition path sampling method (TPS) \cite{Dellago2002} includes only the transition paths, i.e. trajectories connecting the two stable states $A$ and $B$, the TIS method samples all possible trajectories of the ensemble and yields an efficient calculation of rate constants and  free energy profiles. In this approach a collection of non-intersecting interfaces, defined by hypersurfaces along a progress order parameter $\lambda_i$, are introduced between the two stable states. An ensemble of trajectories is sampled for each interface such that a trajectory is accepted in the MC step if it starts and ends in one of the stable states $A$ or $B$, and if it crosses the interface $\lambda_i$. Two kinds of ensembles are sampled with the TIS method,  the {\it forward} ensemble $\mathcal P_{A\lambda_i}$ which includes all trajectories that leave region $A$, and the  {\it backward} ensemble $\mathcal P_{B\lambda_i} $,  which includes the reverse transition, i.e. all trajectories that start in region $B$.

In this work, we sample the TIS ensemble of trajectories using the shooting algorithm,~\cite{Dellago1998} where a slight perturbation of the momenta and/or positions is performed for a configuration randomly selected from a trial path. A new trajectory  is generated by integrating the equations  of motion forward and backward in time from the modified configuration. If the new path begins and ends in either of the stable states $A$ or $B$ and crosses the corresponding interface $\lambda_i$ it is accepted. For an efficient exploration of the trajectory space we further employ exchange moves where trajectories between subsequent interface ensembles are swapped yielding a combination of the TIS algorithm with the replica exchange approach (RETIS).~\cite{VanErp2007,Bolhuis2008}

\subsection{Reweighted path ensemble}

An estimate of the unbiased  path ensemble  can be obtained by reweighting the trajectories of the TIS path ensemble~\cite{Rogal2010}   according to their correct contribution to the Boltzmann factor using the weighted histogram analysis method (WHAM).~\cite{Ferrenberg1989}  
The reweighted path ensemble (RPE)~\cite{Rogal2010}   is given by 
\begin{equation}
\label{eq1}
\mathcal P[\mathbf{x}^L] = c_A \sum_{i=1}^n \mathcal P_{A\lambda_i}[\mathbf{x}^L] W_A[\mathbf{x}^L ]
+ c_B \sum_{i=1}^n \mathcal P_{B\lambda_i}[\mathbf{x}^L] W_B[\mathbf{x}^L ]
\end{equation}
\noindent
where $\mathcal P_{A\lambda_i}[\mathbf{x}^L]$ and $\mathcal P_{B\lambda_i}[\mathbf{x}^L]$ are the TIS forward and backward ensembles of trajectories per interface, $\mathbf{x}^L$ denotes a trajectory or a sequence of $L$ phase space points $\{\mathbf{x}_0,...,\mathbf{x}_L\}$, and $c_A$ and $c_B$ are constant factors obtained from $A-B$ and $B-A$ path histograms to fix the relative weights in the RPE (see Ref.~\onlinecite{Rogal2010}). Here, $W_A[\mathbf{x}^L ]$ and $W_B[\mathbf{x}^L ]$ are the path weights calculated from the crossing probabilities of the paths for each interface by using WHAM.

Once we obtain a complete and unbiased path ensemble other statistical quantities like the free energy and the averaged committor function can be projected on a set of collective variables (CVs or order parameters) $\mathbf{q}=\{q_1(\mathbf{x}),..,q_d(\mathbf{x})\}$: 
\begin{equation}
\label{eq2}
F(\mathbf{q})=-k_{\text{B}}T \ln\rho(\mathbf{q}) + const. 
\end{equation}
\noindent
where $k_{\text{B}}$ is Boltzmann's constant and $\rho(\mathbf{q})$ is the probability density to find a configuration 
$\mathbf{q}$ in the path ensemble $\mathcal P[\mathbf{x}^L]$:
\begin{equation}
\label{eq3}
\rho(\mathbf{q})= C\int\mathcal D \mathbf{x}^L \mathcal P[\mathbf{x}^L] \sum_{k=0}^L \delta(\mathbf{q}(\mathbf{x}_k)-\mathbf{q}) \quad .
\end{equation}
\noindent
Here $\int \mathcal{D}\mathbf{x}^L$ indicates the path integral evaluated in the RPE, $\delta(\mathbf{z}) = \prod_{i=1}^d \delta(z_i)$ is the Dirac delta function and $C$ is a normalization constant.

Similarly, the averaged committor probability can be projected on any set of collective variables~\cite{Bolhuis2011} by histogramming all the trajectories of the RPE that commit to state $B$:~\cite{Bolhuis2011}
\begin{equation}
\label{eq4}
\overline{p}_B(\mathbf{q})= \frac{\int\mathcal D \mathbf{x}^L \mathcal P[\mathbf{x}^L] \mathbbm{1}_B (\mathbf{x}_L) \sum_{k=0}^L \delta(\mathbf{q}(\mathbf{x}_k)-\mathbf{q})}{\int\mathcal D \mathbf{x}^L \mathcal P[\mathbf{x}^L] \sum_{k=0}^L \delta(\mathbf{q}(\mathbf{x}_k)-\mathbf{q})} \quad ,
\end{equation}
where $\mathbbm{1}_B(\mathbf{x}_L)$  is a function that selects trajectories by assigning a value of one to paths that end in state $B$ and zero otherwise.

\subsection{Maximum likelihood estimation}

The RPE holds dynamical information of the committor function that
can be used to analyze the RC of the process by employing 
a maximum likelihood estimation (MLE).~\cite{Husmeier2005, Peters2006}
The committor is defined as the probability $p_B(\mathbf{x})$ that a trajectory 
starting at a phase space point $\mathbf{x}$ will reach state $B$ before coming 
back to state $A$.  Therefore configurations at the transition state yield a committor 
value of $p_B(\mathbf{x})=0.5$, i.e. the trajectory 
has the same probability to reach $A$ or $B$, while $p_B(\mathbf{x})=0$ and 
$p_B(\mathbf{x})=1$ at the stable states $A$ and $B$, respectively. The committor 
$p_B(\mathbf{x})$ is commonly seen as the perfect RC since it provides a 
measure of the progress along the reactive event. However, the committor itself cannot 
be directly compared to experiments  as it lacks a physical interpretation 
of the mechanism. For this reason, a committor $p_B(r)$ is commonly used 
as a model of the perfect RC to evaluate the quality of $r(\mathbf{q})$, a function of a 
set of candidate order parameters $\mathbf{q}=\{q_1(\mathbf{x}),...,q_d(\mathbf{x}) \}$ 
proposed to describe the mechanism. 
Using the MLE approach it can be evaluated which model committor function best fits the data from the path ensemble, allowing a quantitative comparison of different sets of order parameters.~\cite{Peters2006, Peters2007}

In the MLE analysis the committor function $p_B(r)$ is modeled 
with a  $\tanh$ function:
\begin{equation}
\label{eq5}
p_B(r)=\frac{1}{2} \left(1+\tanh(r(\mathbf{x}))\right)
\end{equation}
\noindent
yielding the typical shape along the RC.
In the original work by Peters \emph{et al.}~\cite{Peters2006} the configurations $\mathbf{x}$ of the TPS ensemble are collected near the transition state region,  
for which a linear combination of CVs is proposed to model $r(\mathbf{q})$.
Since in this work we are aiming to capture the entire transition between the stable states $A$ and $B$ we use a 1D string in the $d$-dimensional collective variable space as a non-linear model of $r(\mathbf{x})$ as proposed by Lechner~\emph{et al.}.~\cite{Lechner2010} 
The string consists of  $m$ images  $\mathbf{S}=\{ \mathbf{s}_0,..,\mathbf{s}_m \} $ 
in the CV space $\mathbf{q}$ and is parametrized by a progress parameter $\sigma(\mathbf{S}(\mathbf{x}))\in [0,1]$
where $\sigma(\mathbf{s}_0)=0$ corresponds to state $A$ and $\sigma(\mathbf{s}_m)=1$ corresponds to state $B$. 
The parameter $\sigma$ defines a path-variable that projects the phase space points $\mathbf{x}$ onto the string. 
This is done using a Voronoi construction that assigns each $\mathbf{x}$  to the closest two string images in CV space followed by a  piecewise monotonic 
interpolation to obtain a continues value of $\sigma(\mathbf{S}(\mathbf{q}(\mathbf{x})))$, which is then mapped to the RC by a function $r=f(\sigma)$ (typically a simple monotonic spline function).~\cite{Lechner2010}
The corresponding likelihood is given by
\begin{equation}
\label{eq_likelihood_RPE}
 L = \prod_{\mathbf{x}_i\to B} p_B(r(\mathbf{x}_i))^{W} \prod_{\mathbf{x}_i\to A} \big( 1-p_B(r(\mathbf{x}_i))\big)^{W}
\end{equation}
where $W=W(\mathbf{x}_i)$ are the weights assigned to each phase space configuration of each path $\mathbf{x}^L$ in the RPE, cf. Eq.~\eqref{eq1}.  $\mathbf{x}_i \to B$ denotes the product over all phase space points from trajectories that end in $B$, and $\mathbf{x}_i \to A$ from all paths that end in $A$, respectively. 
Using the expression in Eq.~\eqref{eq_likelihood_RPE} the images of the string and the corresponding $\sigma$ and $r$ are adjusted to maximize the likelihood with a Monte Carlo annealing scheme (described in Ref.~\onlinecite{Lechner2010}). 

For convenience, the likelihood is usually expressed as the logarithm $\ln(L)$.  
The absolute value of the logarithmic likelihood increases with the number of data points (phase space points from the path ensemble) as well as with the number of degrees of freedom in the model (dimension of the CV space, number of string images).
Therefore we cannot directly compare $\ln(L)$ for spaces of different dimensionality.
Instead, the Bayesian information criterion~\cite{Schwarz1978} (BIC) 
is used to determine if there is a significant 
improvement of the likelihood when adding an additional variable. According to the BIC, additional variables benefit the model when  
the following quantity is maximized:
\begin{equation}
\label{bic}
 \ln \mathcal L =\ln(L) - 0.5 n_v \ln(N_d)
\end{equation}
where $N_d=\sum_{\mathbf{x}_i} W(\mathbf{x}_i)$ is the weighted number of data points from the RPE 
and $n_v$ is the number of variables used to describe the model. 
In our case $n_v=(d+1)(m-2)+2$, where $d$ is the dimension of the CV space and $m$ is the number of string images, considering that the
end points of the string are fixed.

\section{Nucleation during solidification in Nickel}
\label{tis_simulations}

\subsection{TIS simulations}
\begin{figure}[tb]
\includegraphics[width=8.0cm,clip=true]{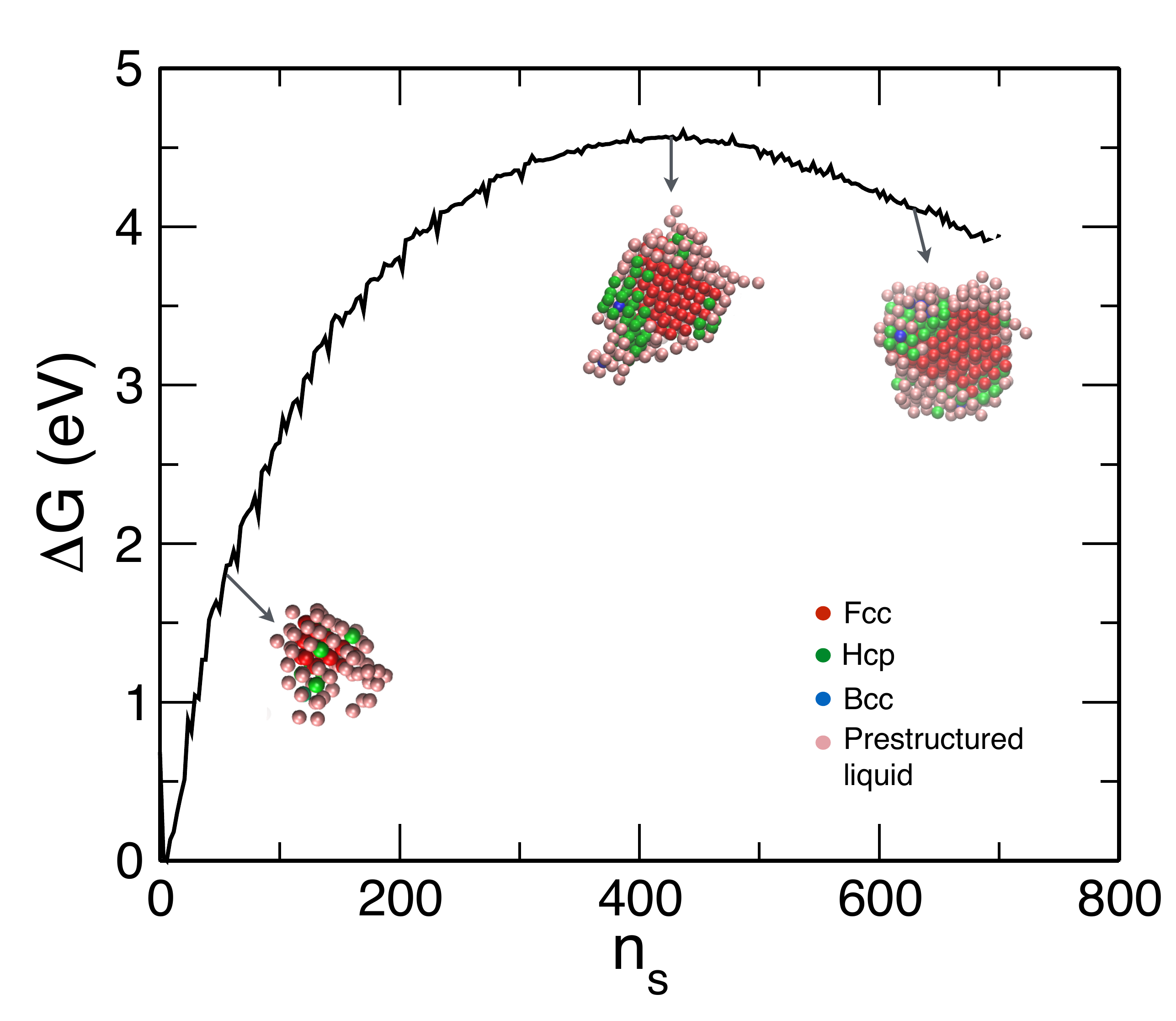}
\caption{\label{fig1} Free energy profile along the largest cluster size $n_s$. Precritical, critical and postcritical representative nuclei  are shown with their structural composition. The growing nucleus exhibits a core composed of fcc (red), surrounded by random hcp (green) and a nucleus surface composed predominately of pre-structured liquid (light brown) with higher correlation than the liquid but less symmetry than a crystal. 
}
\end{figure}

For the reaction coordinate analysis of nucleation during solidification in homogeneous Ni,
we perform RETIS simulations at a moderate undercooling (cf. Ref.~\onlinecite{DLeines2017}).
The ensemble of trajectories is sampled between the liquid and solid phases 
in a system of 8788 Ni atoms.  We employ the embedded atom 
method (EAM) potential by Foiles \emph{et al.}~\cite{Foiles86} to model the interaction between the Ni atoms. 
All molecular dynamics (MD) simulations are performed at constant pressure $P=0$~bars and 
temperature $T=1370$~K (NPT ensemble), which corresponds to an undercooling of $\Delta T/ T_m \sim 20 \%$.
The MD simulations were performed with the package LAMMPS,~\cite{Plimpton1995} using a time step of 
$2$~fs and minimum image periodic boundary conditions in all directions.
The LAMMPS code was combined with a python wrapper to carry out the RETIS simulations.
In the RETIS simulations, we performed 45\% shooting moves, 45\% exchange moves
and 10\% exchange moves between forward and backward ensembles for the MC sampling of trajectories. 
A total of 2500 MC moves were performed with paths being collected in the ensemble after 5 decorrelation steps yielding 500 trajectories for each interface.
The amplitude of the velocity perturbation in the shooting moves was adjusted such that an acceptance ratio larger than $30\%$ is obtained. The time interval between
slices stored along a path was set to 0.2~ps. The positions of the interfaces were set such that
the crossing probabilities overlap by at least $10\%$. We employ the largest solid
cluster size $n_s$ as an order parameter to define the interfaces of the ensemble (see Sec.~\ref{opsection} 
for a definition of $n_s$).
The first interface $n_s=20$ defines the boundary of the undercooled liquid region, and
the last interface the boundary of the solid state region with $n_s = 700$.
All configurations with $n_s > 700$ are fully committed to the bulk phase.

At the chosen undercooling  the free energy barrier obtained from the RPE and projected onto the largest cluster size $n_s$ 
is  $\Delta G^* = 4.5$~eV (see Fig.~\ref{fig1}), with an associated timescale of hundreds of seconds. 
Our previous simulations show that the free energy barriers for undercoolings in the range of $20\%-25\%$ are in close quantitative 
agreement with CNT and experiments.~\cite{DLeines2017} 
However, the analysis of the RPE in Ref.~\onlinecite{DLeines2017} reveals that the \emph{mechanism} of nucleation 
in Ni differs from the one predicted by CNT:   the solid clusters are mostly non-spherical, exhibit random fcc-hcp stacking, and 
have a diffusive  solid-liquid interface. As illustrated in Fig.~\ref{fig1}, the growing nuclei are composed of  fcc atoms 
 in the  core region surrounded by random hcp (stacking faults) and a pre-structured 
liquid cloud that forms the diffusive surface between the liquid and the crystal.  Bcc coordinated atoms are
rarely found in the growing nuclei. The nucleation process was also found to initiate in long-lived regions of 
pre-structured liquid characterized by a higher correlation than the liquid but less symmetry than 
the crystal phases (a mesocrystal phase) followed by a subsequent emergence of  the crystal phase 
within the core of the pre-structured clusters.~\cite{DLeines2017}
To quantify the role of the different crystal phases in the growing clusters and mesocrystal regions
observed during the nucleation, we perform a MLE analysis of various structural CVs to identify the optimal RC that describes the nucleation process.

\subsection{Order parameters to study nucleation }
\label{opsection}

We now introduce a set of order parameters  that serve as candidates for the RC of the crystallization mechanism
during homogeneous nucleation in Ni.  A standard approach to distinguish between solid-like and liquid-like particles
was introduced by ten Wolde and Frenkel, based on the Steinhardt  
bond order parameters.~\cite{Steinhardt1983, Auer2005} In this method, the
structural correlation in the neighborhood of each particle is obtained using
a local criterium of solidity.  A particle $i$ is connected to a neighbor particle $j$ in a solid bond 
if the correlation $s_{ij}  =  \sum_{m=-6}^{6} q_{6m}(i)q_{6m}^*(j)> 0.5$, where 
$q_{6m}$ are the complex vectors based on the spherical harmonics.~\cite{Steinhardt1983} 
If a particle has between $6-8$ solid connections, it is identified 
as solid.  To improve the solidity distinction criterium at 
the interface of the solid clusters, another parameter is defined as the average of
the correlation over the nearest neighbors $\langle s_{ij} \rangle =1/N_\text{nn}\sum s_{ij}$. This
parameter includes a measure of the disorder surrounding a particle $i$:  if 
 $\langle s_{ij}\rangle > 0.6$, particle $i$ is considered as solid. 
These two criteria together with a clustering algorithm of 
nearest neighbors are used to define the largest solid cluster size $n_s$.

The crystal structures of the solid particles in the cluster are discriminated 
using the averaged   Steinhardt  bond order parameters $\bar{q}_4$ and $\bar{q}_6$.~\cite{Lechner2008}
These parameters allow a local distinction of the crystal structure of a particle, as well as the liquid-like particles. 
Since the reference histograms in the $\bar{q}_4-\bar{q}_6$ plane of the perfect crystal structures hcp, 
bcc, fcc, and of the liquid show little overlap, we can assign a crystal structure to each particle. 
For this the $\bar{q}_4, \bar{q}_6$ values are calculated for a particle and the corresponding probabilities of the various crystal structures are evaluated from the reference distributions in the $\bar{q}_4-\bar{q}_6$-plane.  The structure with the largest probability is assigned to the particle.
If all  probabilities are smaller than $10^{-5}$  the particle is labeled  as  \emph{undefined}.

The averaged bond order parameters $\bar{q}_4-\bar{q}_6$ provide another criterium of solidity
where the particles are identified as crystalline if the liquid probability vanishes.~\cite{Lechner2011}
Using these  parameters we define other structural collective variables for the RC model. 
We use the solidity definition of ten Wolde~\emph{et al.}  together with the averaged bond order parameters 
to define the largest \emph{crystalline} cluster size $n_c$, which includes only particles with a crystalline structure,
 e.g. hcp, bcc, fcc. 
Other variables analyzed within the MLE method are the number of fcc, hcp, bcc particles 
$n_\text{fcc}$, $n_\text{bcc}$ and $n_\text{hcp}$ in the largest cluster. We further define
 the number of fcc, hcp, bcc atoms  in the entire system $N_\text{fcc}$, $N_\text{bcc}$ and $N_\text{hcp}$.

To investigate the role of the formation of preordered regions in the RC, we determine the number of pre-structured liquid particles in the largest cluster $n_\text{pl}$ and in the total system $N_\text{pl}$. The pre-structured
liquid particles are identified using the histograms of fcc, hcp, bcc, and liquid structures 
on the $\bar{q}_4(i)-\bar{q}_6(i)$ map. A particle is assigned to be pre-structured liquid if the solidity criterion of ten Wolde~\emph{et al.} is fulfilled, but all crystal probabilities are smaller than $10^{-5}$,  and the liquid probability vanishes. 
These regions therefore have less symmetry than the crystal structures but higher orientational 
order than the liquid, and their $\bar{q}_6$ and $\bar{q}_4$ values lie in between the liquid and crystal regions.~\cite{DLeines2017}


\section{Optimal reaction coordinates}
\label{RCanalysis}
\subsection{MLE of single order parameters}
\label{sec:MLE1D}

\begin{figure*}[tb]
\includegraphics[width=17.0cm,clip=true]{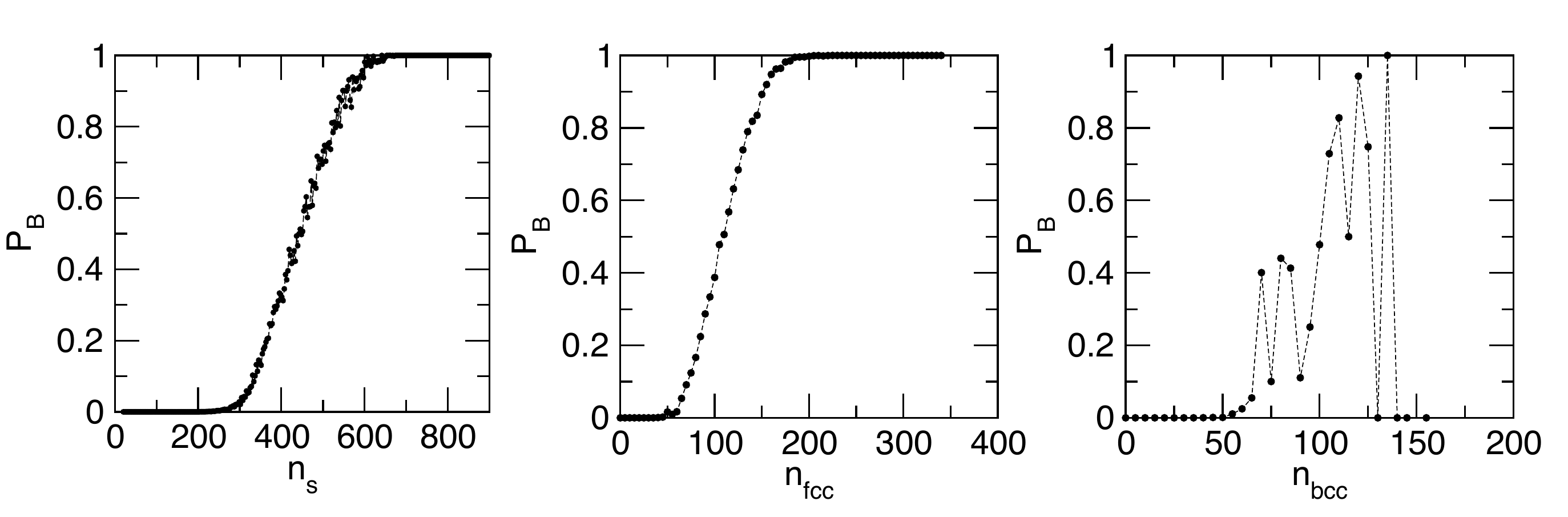} 
\caption{Averaged committor function $p_B$ obtained from the RPE and projected onto different CVs.  From left to right, we show $p_B$ projected on CVs selected from the highest to the lowest rank in the maximum likelihood analysis: the number of particles in the largest solid cluster $n_s$, the fraction of fcc particles in the largest cluster  $n_\text{fcc}$ and the fraction of bcc particles in the largest cluster $n_\text{bcc}$.  \label{comm_1d} 
}
\end{figure*}

\begin{table}
\caption{Single order parameters ranked by their BIC $\ln \mathcal L$ and normalized by the BIC of the largest 
cluster size $\ln \mathcal{L}_{n_s}=-63.9$. The likelihood is calculated from 5 independent MC annealing runs and the relative error $\epsilon$ is given in percentage by the standard deviation of the  data.
}
\begin{ruledtabular}
\begin{tabular}{llccccc}
\label{table1-lnL}
   Rank & Order parameter & $\ln \mathcal L_{n_s} / \ln \mathcal L$ & $\epsilon$ (\%)  \\ \hline 
 1 & $n_s$ & $1.00$  &  $0.02$ \\ 
  2 & $n_c$ &$0.8646$ & $0.02$ \\
  3 & $n_\text{pl}$ & 0.7955 & $0.02$ \\
  4 & $n_\text{fcc}$ &0.7909 & $0.04$ \\
  5 & $N_\text{fcc}$ & 0.7916 & $0.03$ \\
  6 & $N_\text{pl}$ & 0.7415 & $0.02$ \\
  7 & $N_\text{hcp}$ & 0.6082 & $0.03$ \\
  8 & $n_\text{hcp}$ & 0.6083 & $0.02$  \\
  9 & $n_\text{bcc}$ & 0.0600 & $0.02$ \\
  10 & $N_\text{bcc}$ & 0.0602 & $0.02$  \\
\end{tabular}
\end{ruledtabular}
\end{table}

We first maximize the BIC $\ln \mathcal{L}$ in Eq.~\eqref{bic} 
using the MLE approach for each  order parameters  described in section~\ref{opsection}  to find the best one-dimensional description of the nucleation mechanism. 
We included configurations from all slices of 300 paths  for each interface, a total of 699 558 points,  as data points from the RPE in the MLE.
The maximization of the likelihood is performed using a Monte Carlo annealing method~\cite{Lechner2010} for different numbers of string images. In 1D CV spaces the string images are fixed and a MC optimization of the $\sigma$ to  $r$ mapping function is performed to maximize the likelihood.  The function $f(\sigma)$ is initially defined by the $M$ images of the string as a piecewise monotonically increasing function. A random displacement of the progress parameter associated with the string images $\sigma(\mathbf{S})$ is performed such that the trial function $f'(\sigma)$ remains monotonically increasing. If the BIC $\ln \mathcal{L}$ 
 increases the trial function is accepted and otherwise rejected.  The mapping functions were optimized for $2 \times 10^5$ MC steps in total.   For all CVs a maximum $\ln \mathcal{L}$ was obtained for $M=6$ images along the strings.

In Tab.~\ref{table1-lnL} the maximum BIC values normalized by the BIC of the largest cluster size, $\ln \mathcal L_{n_s}=-63.90$, are listed for all tested order parameters.
The size of the largest cluster, $n_s$, occurs to be the best approximation to the RC.
It  enhances the description of the nucleation process by $\sim 14\%$  in comparison to the number of particles in the \emph{crystalline} cluster, $n_c$, that ranks  second,  and is $\sim 20\%$ better than the number of fcc particles in the cluster, $n_\text{fcc}$. 
This contradicts the capillarity assumption within CNT:  the bulk phase of Ni is fcc, and therefore according to CNT the order parameter $n_\text{fcc}$ would be considered as the sole RC of the process. 
However, the MLE analysis reveals that it is by far not the best representation of the RC.  
Interestingly, the number of fcc and pre-structured liquid particles in the largest cluster have similar likelihoods, while the rest of the order parameters perform poorly as approximations to the RC, especially $n_\text{hcp}$, $N_\text{hcp}$, $n_\text{bcc}$, and $N_\text{bcc}$. From the structural analysis of the RPE we know  that the average composition of the growing nucleus consists of fcc particles surrounded by random hcp (stacking faults) that emerge within the core of the pre-structured liquid cloud.~\cite{DLeines2017} The average amount of bcc within the growing solid cluster is negligible along the transition pathways. We therefore expect that $n_\text{bcc}$ and $n_\text{hcp}$ perform rather poorly as RC candidates of the nucleation process, as corroborated by the 
MLE analysis, cf. Tab.~\ref{table1-lnL}.
This is further visualized in Fig.~\ref{comm_1d}, where the averaged committor (Eq.~\eqref{eq4}) has been projected onto different  CVs.  As the committor is the perfect RC, the best CVs identified in the MLE analysis  must approximate the committor function closely.
Fig.~\ref{comm_1d} shows that  $n_s$  and $n_\text{fcc}$ are closely correlated with a smooth averaged committor function $p_B(n_s)$ and $p_B(n_\text{fcc})$, while other variables of very low likelihood such as $n_\text{bcc}$ show a scattered and poor approximation to the committor function.

Although the size of the fcc crystal nucleus $n_\text{fcc}$ shows a qualitatively good correlation with the averaged committor,  the MLE analysis in Tab.~\ref{table1-lnL}  reveals that quantitatively $n_\text{fcc}$ is not the CV of maximum likelihood and therefore is not the order parameter that fully describes the process. 
Our analysis further shows that including other crystalline phases in the description of the solid cluster, 
as represented by $n_c$, 
improves  the RC model only slightly by $\sim 7\% $ compared to $n_\text{fcc}$. 
Therefore, the improvement of the RC model when considering $n_s$ comes mainly from the inclusion of the  pre-structured liquid region in the description of the solid clusters.

\begin{figure}[tb]
\includegraphics[width=8.0cm,clip=true]{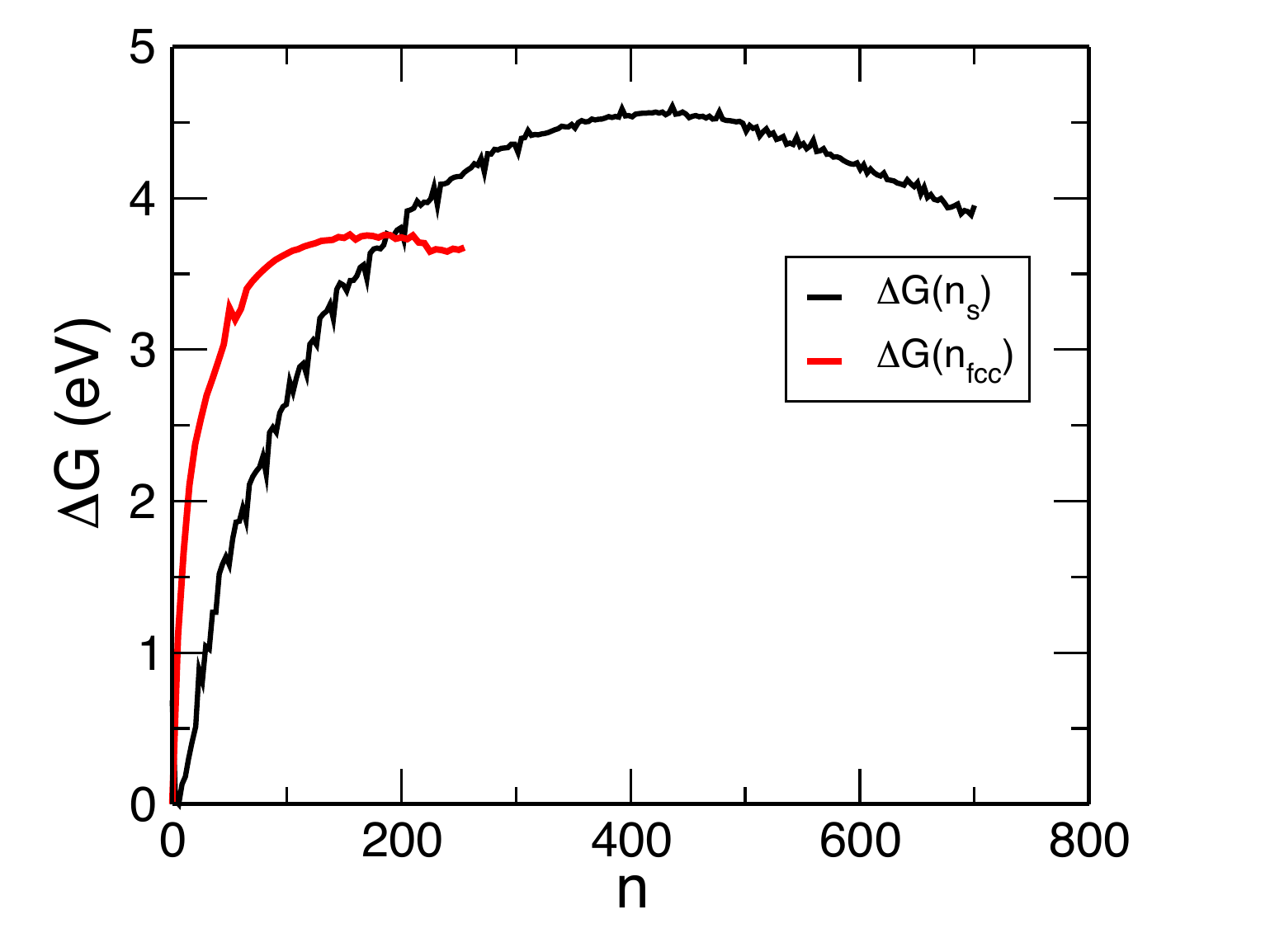} 
\caption{Free energy profiles (left)  obtained from the RPE and projected on the largest nucleus size $n_s$ (black) and on the size of the fcc crystalline core $n_\text{fcc}$ (red).  \label{fig3} 
}
\end{figure}

Fig.~\ref{fig3} shows the free energy profiles projected from the RPE onto $n_s$ and $n_\text{fcc}$. The  free energy projected onto $n_\text{fcc}$, the order parameter assumed by CNT, yields a barrier $\Delta G^*(n_\text{fcc})=3.7\pm 0.05$ eV which is $~17\%$ lower than the barrier estimated for the projection onto $n_s$. Moreover, the critical nucleus size calculated from the RPE using the averaged committor at $p_B=0.5$ and projected onto $n_s$ and $n_{fcc}$ 
indicates a difference in size of around 318 particles.  This large difference in the barrier and critical nucleus size can be understood from the structural analysis of the nucleation mechanism in Ni.  The pre-structured liquid cloud is a region of hidden order that initially emerges within the liquid and a subsequent nucleation of crystallites is enhanced by these mesocrystalline clouds which act as seeds or precursors of the crystallization. The difference in the free energy barrier is the result of this hidden order of the pre-structured region within the liquid, which is not considered in the free energy projection onto the fcc cluster $n_\text{fcc}$. Solid clusters that are composed of more than 90\%  pre-structured liquid emerge initially within the liquid up to cluster sizes of 50 - 100 particles with a corresponding increase in the free energy by $\sim 1.0$ eV (which is missing in the projection $\Delta G^*(n_\text{fcc})$). As the crystallites grow within the core of the clusters, the free energy projection onto $n_s$ includes a contribution from a diffusive interface region of pre-structured liquid that surrounds the crystal core. Therefore, the higher free energy  as a function of $n_s$ can be attributed to the initial fluctuations of high orientational order to form the mesocrystal seeds within the liquid, and to the interfacial free energy contribution of the pre-structured cloud during the crystal phase growth. The good agreement of the free energy barrier  $\Delta G^*(n_s)$ with experiments~\cite{Bokeloh2011}  as well as the MLE analysis 
show that  the pre-structured liquid plays a prominent role in the description of the nucleation process.

\subsection{MLE in two-dimensional space}
\label{sec:MLE2D}

As a next step we  investigate if the description of the  reaction coordinate can be further improved in a 2D CV space.  For this we define  combinations of $n_s$ and $n_c$ with all other order parameters,  $(n_s, q_i)$ and $(n_c, q_i)$. 
To maximize the likelihood in each of the 2D spaces a string with $M$ images is optimized by two separate MC moves:
one for  the $\sigma$ to $r$ mapping (as done in the one-dimensional space) and a second for the position of the string. An image of the string is randomly selected and displaced by a small distance. If the likelihood increases for the trial string we accept the move and otherwise reject it. 
We have also tested different numbers of images along the string and found the maximum BIC $\ln \mathcal{L}$ 
for $M=3$ images. The initial 
and the final images of the strings are kept fixed at the position of the stable states obtained from the 2D free energy projections. 
The logarithmic likelihoods of each 2D space were optimized for 1000 string displacements and $2\times10^5$ moves of the mapping function.

\begin{table}
\caption{
Two-dimensional combinations of order parameters ranked by their BIC $\ln \mathcal L(q_i,q_j)$ and normalized by the BIC of the largest cluster size $\ln \mathcal L_{n_s}$. The likelihood is calculated from 5 independent runs and the relative error $\epsilon$  is given in percentage by the standard deviation of the data.
}
\begin{ruledtabular}
\begin{tabular}{llccccc}
\label{table2-lnL}

   Rank & Order parameter & $\ln \mathcal L_{n_s} / \ln \mathcal L$ & $\epsilon$ (\%)  \\ \hline
   1 & $(n_s,n_{pl})$ & 0.946 &0.7   \ \\  
   2 & $(n_s,n_\text{c})$ & 0.942 &0.9  \ \\
   3 & $(n_s,n_\text{fcc})$ & 0.944 &0.1  \ \\ 
   4 & $(n_c,n_\text{pl})$ & 0.940 &0.3\ \\
   5 & $(n_c,n_\text{fcc})$ & 0.78 &3.0\ \\     
\end{tabular}
\end{ruledtabular}
\end{table}

Tab.~\ref{table2-lnL} shows the maximum BIC for the 2D CV spaces. According to the Bayesian information criterion,~\cite{Schwarz1978} the information gain of the likelihood with an additional variable in the model should be at least  $1/2 \ln N_d$ to be considered as an improvement. Therefore, additional complexity in the model improves the RC description only  if the BIC $\ln \mathcal L$ increases. Here, the BICs $\ln \mathcal L(q_i, q_j)$ of the string models in 2D are compared to the BIC of the best CV candidate in one dimension $n_s$. If $\ln \mathcal L_{n_s}/ \ln \mathcal L(q_i, q_j) > 1.0 $, there is a significant gain of information in the RC model.
None of the pairs that include $n_s$ or $n_c$ show a relevant gain in information compared to the likelihood of the best one-dimensional RC, $n_s$. 
The combinations including $n_s$  always exhibit a larger likelihood than the ones with $n_c$  which emphasizes the importance of $n_s$ as part of the RC.
The MLE analysis shows that the order parameter $n_s$ can be considered as a nearly complete reaction coordinate. This could explain the good agreement of the nucleation barriers $\Delta G^* (n_s)$ with experiments.~\cite{Bokeloh2011}

\begin{figure*}[tbh!]
\includegraphics[width=15.0cm,clip=true]{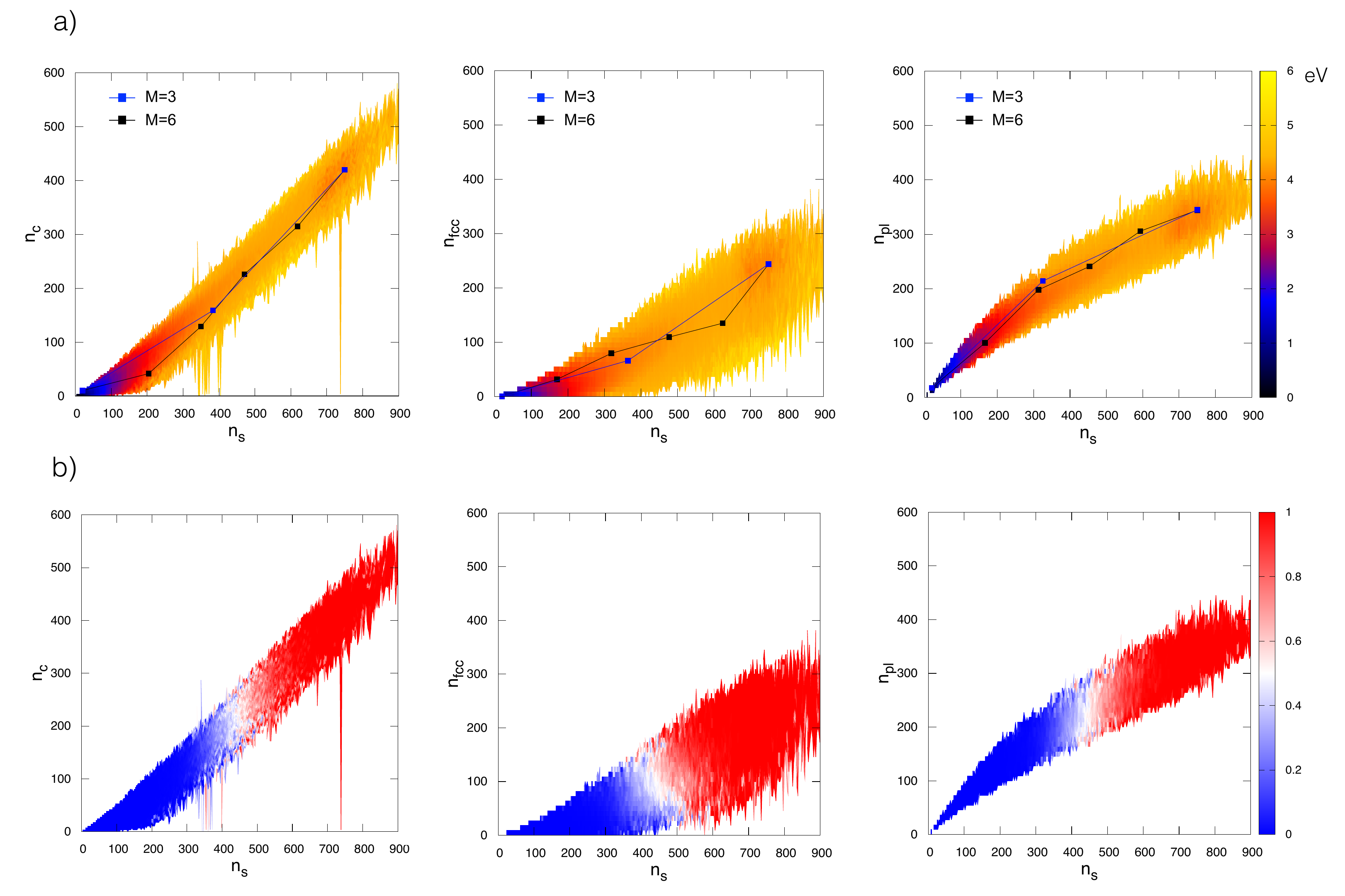}
\caption{\label{fig4} (a) Optimized strings obtained from the MLE analysis and free energy projections in the $(n_s, n_c)$, $(n_s, n_\text{fcc})$, and $(n_s, n_\text{pl})$ space. The final strings of M=3 and M=6 images are shown (black and blue lines). 
The projection of the free energies includes 300 paths per interface from the RPE. 
(b) 2D averaged committors for $(n_s, n_c)$, $(n_s, n_\text{fcc})$, and $(n_s, n_\text{pl})$ obtained from the RPE.
}
\end{figure*}

In  Fig.~\ref{fig4}~a) the averaged optimized strings are shown on the 2D free energy landscapes obtained from the RPE. 
We particularly focus on the combination of $n_s$ with $n_c$, $n_\text{fcc}$, and $n_\text{pl}$ to obtain further insight into the role of the different order parameters during the nucleation process.
The left graph in Fig.~\ref{fig4}~a) shows the projection onto the largest cluster size $n_s$  and the crystal core $n_c$. The optimized string  in the narrow reaction channel demonstrates that $(n_s, n_c)$ are highly correlated and increase linearly for most parts along the centre of the free energy valley. This correlation explains why $n_s$ and $n_c$ are the two candidates with the highest likelihoods in the one-dimensional MLE analysis.  
However, for small cluster sizes (up to $n_s \sim 100 $ and $\Delta G \sim  1.5 $~eV), the number of crystalline particles $n_c$ is nearly zero, 
confirming that the clusters are initially composed of pre-structured liquid followed by a subsequent nucleation of the crystalline phase $n_c$.  A projection of the free energy landscape onto the single variable $n_c$ cannot capture the initial formation of pre-structured liquid clusters leading to a poor model of the RC in this region. This is consistent with our results that in 1D the BIC  $\ln \mathcal{L}_{n_c}$ is lower than $\ln \mathcal{L}_{n_s}$. 
Similarly, the committor projection in the left graph of Fig.~\ref{fig4}~b) clearly shows that the distribution is rather broad along $n_c$, whereas a projection onto $n_s$ captures a well-defined transition state region (narrow white region with $p_B = 0.5$).  This illustrates why there is no significant information gain in this 2D projection as compared to $n_s$.

In the middle graph of Fig.~\ref{fig4}~a) the optimized string in the $(n_s, n_\text{fcc})$ space is shown.  Comparable to the evolution of $n_c$, the number of fcc particles, $n_\text{fcc}$, hardly increases along the optimized string up to cluster sizes of $n_s \sim 200$ associated with free energies of $\Delta G \sim 2$~eV.  
This illustrates that also $n_\text{fcc}$ cannot capture the initial formation of the pre-structured liquid region at small cluster sizes, missing a relevant contribution to the free energy barrier as shown in Fig.~\ref{fig3}.
The committor projection shown in the middle graph of Fig.~\ref{fig4}~b) is even broader along $n_\text{fcc}$ than along $n_c$ which is consistent with the MLE ranking of the CVs in 1D.  
One should keep in mind that the 1D committor projections shown in Fig.~\ref{comm_1d} represent and \emph{averaged} committor which qualitatively appears reasonable for $n_\text{fcc}$, but quantitatively $n_\text{fcc}$ does not represent the best RC.
Interestingly, the transition state region is not entirely parallel to $n_\text{fcc}$, which indicates that the critical nucleus size in terms of $n_s$ depends on the composition:  the larger the amount of fcc particles in the growing nucleus the smaller the critical nucleus size.

The free energy projection and optimized string in the $(n_s, n_\text{pl})$ space is shown in the right graph of Fig.~\ref{fig4}~a).  For  small clusters sizes ($n_s < 100$) the energy landscape is narrow, $n_s$ and $n_\text{pl}$ are highly correlated, and the optimized string is practically linear.  For larger cluster sizes, however, $n_s$ and $n_\text{pl}$ exhibit a nonlinear behavior  where $n_s$ increases faster than $n_\text{pl}$. 
A fit to the data for small cluster sizes $n_s < 100$ yields $n_\text{pl}=0.9 n_{s}^{0.9}$, confirming the linear relationship between the two CVs.  This demonstrates that the first step in the nucleation mechanism is the formation of pre-structured liquid clusters.  For larger clusters $n_s > 100$, we obtain $n_{pl}=3.1 n_{s}^{0.7}$ which is very close to the surface-volume ratio of CNT $n_\text{surf} \propto n_\text{vol}^{0.68}$. 
The committor in the right graph of Fig.~\ref{fig4}~b) exhibits again a fairly narrow transition state region with respect to $n_s$, whereas the distribution along $n_\text{pl}$ is rather broad, illustrating that a projection onto  $n_s$ can well describe the transition state.
The single parameter $n_\text{pl}$ lacks, however, information on the formation of the crystal core in the cluster and is not the best single descriptor of the nucleation transition, as shown in Sec.~\ref{sec:MLE1D}. 

Our analysis indicates that the nucleation mechanism consists of  an initial formation of pre-structured liquid clusters followed by the growth of crystallites embedded in a diffusive surface, mostly composed of pre-structured particles. This is in agreement with other studies of crystallization in hard spheres (comparable model systems to metals~\cite{Herlach2016}) and soft core colloidal models \cite{Lechner2011,PhysRevLett.105.025701,Kawasaki2011,Russo2012,Russo2012b, PhysRevLett.105.025701} where a preordered liquid region acts as precursor of the nucleation process and plays a key role in the surface description of the RC.


\subsection{Role of pre-structured liquid clusters as precursors for nucleation}

\begin{figure}
\includegraphics[width=6.5cm,clip=true]{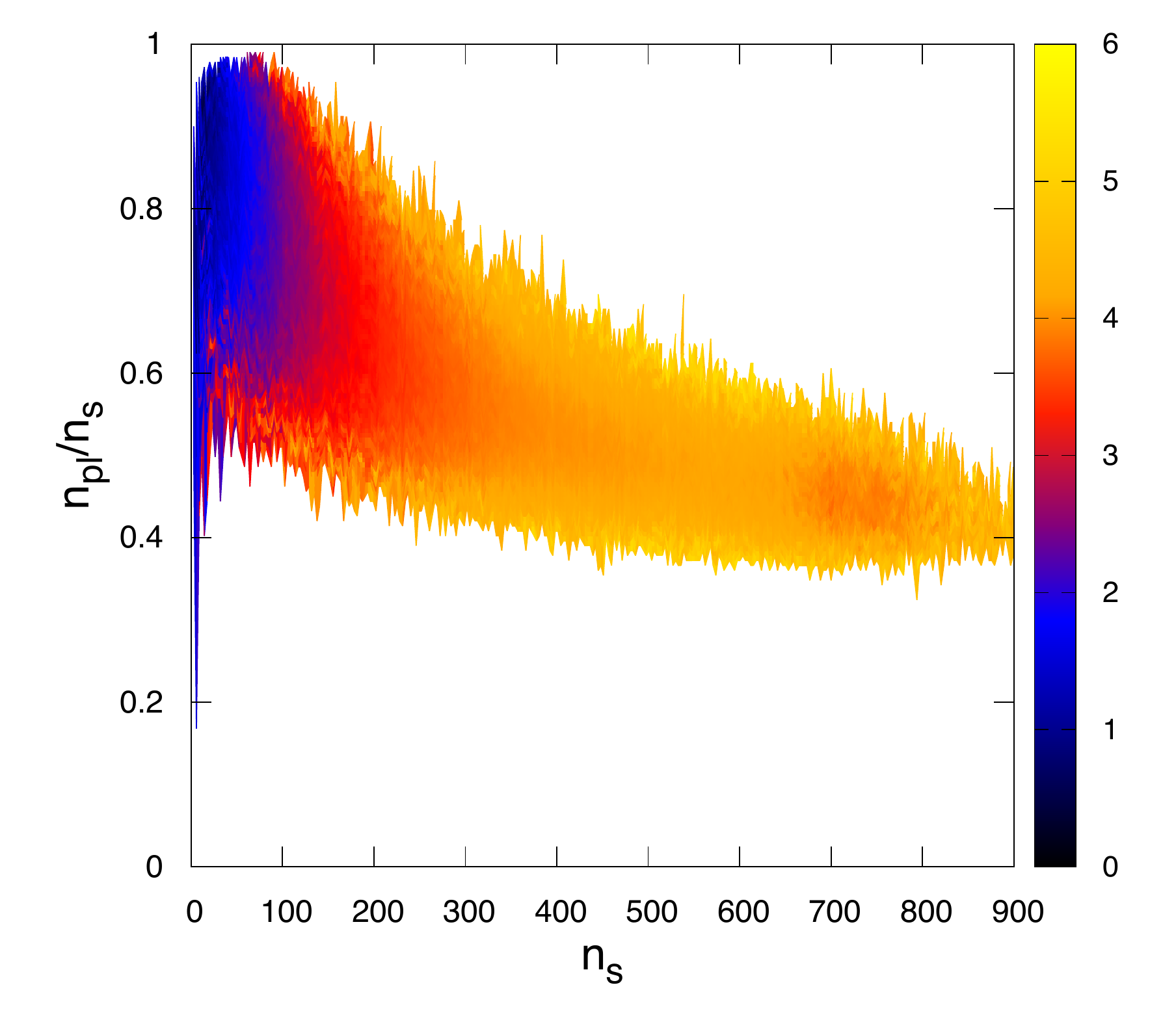}
\caption{\label{fig:fes_npl}  Two dimensional projection of the free energy onto the fraction of pre-structured liquid particles in the largest solid cluster ($n_\text{pl}/n_s$) and the largest cluster size $n_s$.  The free energy  is projected from the RPE including all the configurations of 350 paths per interface. The color bar indicates the magnitude of the  free energy  $\Delta G (n_s,n_\text{pl}/n_s)$ in eV. 
}
\end{figure}
Fig.~\ref{fig:fes_npl} shows a projection of the  free energy as a function of the fraction of pre-structured liquid particles in the largest cluster $n_\text{pl}/n_s$ and the size of largest solid cluster $n_s$. The stable state at high values of $n_\text{pl}/n_s=0.9$  evidences the initial formation of the pre-structured liquid cluster that precedes the growth of the crystal nucleus. Solid clusters of 50-60 particles are mostly composed of pre-structured liquid ($n_\text{pl}/n_s \sim 0.9$) and associated with free energies up to  $\Delta G \sim 1.5$~eV, which constitutes a significant part of the nucleation barrier. As crystalline particles emerge  within the core of the  nucleus, the fraction of pre-structured liquid particles in the cluster decreases.  
But even for large cluster sizes of $\sim 700$ particles the fraction of pre-structured liquid remains $n_\text{pl}/n_s \sim 0.5$.
This mesocrystal region thus exhibits a long lifetime and decays rather slowly even beyond the transition state region, i.e. during crystal growth.

The free energy projection $\Delta G(n_s,n_\text{pl}/n_s)$ shown in Fig.~\ref{fig:fes_npl} corroborates the two-step nucleation mechanism discussed in the previous section and emphasizes the importance of the pre-structured liquid as precursor.  
To further analyze the structure of the preordered region we employ a polyhedron analysis for structure identification via topological fingerprints.~\cite{Schablitzki2013}  The distribution of coordination polyhedra in the preordered region clearly deviates from the one in the liquid.  12-fold coordinated atoms are dominant with polyhedra that resemble fcc-hcp like symmetries, closely related to polyhedra found in fcc bulk (including thermal vibrations).  Here, the pre-structured liquid cloud does not only act as a seed for crystal nucleation, but also predetermines which polymorph nucleates which in turn defines the final bulk structure.

The role of $n_\text{pl}$ in the enhancement of the reaction coordinate together with the long lifetime of preordered regions during the structural evolution of the nucleus evidences that the pre-structured liquid is not a trivial intermediate ordering step between the solid and the liquid, but rather a mesocrystal phase that acts as precursor during crystal nucleation.

\section{Conclusions}
\label{conclusion}
Using the MLE approach together with the data from the reweighted path ensemble we are able to quantitatively evaluate the quality of different CVs as reaction coordinates. 
For the nucleation in Ni the RC is best described by the structural order parameter $n_s$, which includes
information about the crystal phases in the solid cluster together with the surrounding pre-structured liquid region.  This pre-structured
liquid cloud significantly enhances the description of the  RC for nucleation compared to other order parameters. Qualitatively, 
the general aspects of  homogeneous nucleation  in Ni are captured  by CNT, i.e. there is a well-defined nucleation barrier that can 
be described by a single order parameter corresponding to the largest nucleus size.  However, quantitively  
we have demonstrated that the parameter $n_\text{fcc}$ is by far not the best descriptor of the nucleation process in Ni as assumed 
by CNT.  The initial formation of the pre-structured liquid region and the nucleus surface  are poorly modeled by  $n_\text{fcc}$  
resulting in an underestimation of the nucleation barrier and the critical nucleus size. 

Analyzing the nucleation process along the best RC $n_s$ reveals that the nucleation mechanism exhibits 
a non-classical behavior where pre-structured clusters emerge within the liquid preceding the crystal nucleation. These nucleation 
precursors are clusters composed of an intermediate structure between liquid and crystal that resembles defective hcp-fcc like structures. 
The preordered region also dominates the diffusive surface of the growing crystal clusters described by $n_s$.
We find that the surface-volume ratio predicted by CNT is in agreement with our results when including the pre-structured cloud 
in the description of the solid cluster. This can explain the good agreement of the temperature dependence of the calculated free energy barriers 
with CNT~\cite{DLeines2017, Bokeloh2011} and provides a classical re-interpretation of a non-classical mechanism. 

The emergence of a hidden long-lived preordered region in the liquid and the subsequent emergence of crystallites within 
the cluster core demonstrates that the nucleation in Ni is a two-step process.  A careful analysis of the structural  composition of all solid clusters in the system shows that  the crystal phases (fcc and random hcp) always grow embedded within the center of the solid clusters. Here the spatial correlation within the clusters illustrates how the pre-structured liquid region is a seed for the crystal nucleation. The role of $n_{s}$ in the enhancement of the RC together with the large and long-lived $n_{pl}$ contribution to the nucleus evidences that the pre-structured liquid is not a trivial intermediate ordering step between the solid and the liquid, but rather a mesocrystal phase that acts as precursor of the crystal nucleation.


\section*{Acknowledgements}
We acknowledge financial support by the Mexican National Council for Science and Technology (CONACYT) 
through project 232090 and by the German Research Foundation (DFG) through
project RO~3073/6-1.
 

\begin{thebibliography}{41}%
\makeatletter
\providecommand \@ifxundefined [1]{%
 \@ifx{#1\undefined}
}%
\providecommand \@ifnum [1]{%
 \ifnum #1\expandafter \@firstoftwo
 \else \expandafter \@secondoftwo
 \fi
}%
\providecommand \@ifx [1]{%
 \ifx #1\expandafter \@firstoftwo
 \else \expandafter \@secondoftwo
 \fi
}%
\providecommand \natexlab [1]{#1}%
\providecommand \enquote  [1]{``#1''}%
\providecommand \bibnamefont  [1]{#1}%
\providecommand \bibfnamefont [1]{#1}%
\providecommand \citenamefont [1]{#1}%
\providecommand \href@noop [0]{\@secondoftwo}%
\providecommand \href [0]{\begingroup \@sanitize@url \@href}%
\providecommand \@href[1]{\@@startlink{#1}\@@href}%
\providecommand \@@href[1]{\endgroup#1\@@endlink}%
\providecommand \@sanitize@url [0]{\catcode `\\12\catcode `\$12\catcode
  `\&12\catcode `\#12\catcode `\^12\catcode `\_12\catcode `\%12\relax}%
\providecommand \@@startlink[1]{}%
\providecommand \@@endlink[0]{}%
\providecommand \url  [0]{\begingroup\@sanitize@url \@url }%
\providecommand \@url [1]{\endgroup\@href {#1}{\urlprefix }}%
\providecommand \urlprefix  [0]{URL }%
\providecommand \Eprint [0]{\href }%
\providecommand \doibase [0]{http://dx.doi.org/}%
\providecommand \selectlanguage [0]{\@gobble}%
\providecommand \bibinfo  [0]{\@secondoftwo}%
\providecommand \bibfield  [0]{\@secondoftwo}%
\providecommand \translation [1]{[#1]}%
\providecommand \BibitemOpen [0]{}%
\providecommand \bibitemStop [0]{}%
\providecommand \bibitemNoStop [0]{.\EOS\space}%
\providecommand \EOS [0]{\spacefactor3000\relax}%
\providecommand \BibitemShut  [1]{\csname bibitem#1\endcsname}%
\let\auto@bib@innerbib\@empty
\bibitem [{\citenamefont {Sosso}\ \emph {et~al.}(2016)\citenamefont {Sosso},
  \citenamefont {Chen}, \citenamefont {Cox}, \citenamefont {Fitzner},
  \citenamefont {Pedevilla}, \citenamefont {Zen},\ and\ \citenamefont
  {Michaelides}}]{Sosso2016}%
  \BibitemOpen
  \bibfield  {author} {\bibinfo {author} {\bibfnamefont {G.~C.}\ \bibnamefont
  {Sosso}}, \bibinfo {author} {\bibfnamefont {J.}~\bibnamefont {Chen}},
  \bibinfo {author} {\bibfnamefont {S.~J.}\ \bibnamefont {Cox}}, \bibinfo
  {author} {\bibfnamefont {M.}~\bibnamefont {Fitzner}}, \bibinfo {author}
  {\bibfnamefont {P.}~\bibnamefont {Pedevilla}}, \bibinfo {author}
  {\bibfnamefont {A.}~\bibnamefont {Zen}}, \ and\ \bibinfo {author}
  {\bibfnamefont {A.}~\bibnamefont {Michaelides}},\ }\href {\doibase
  10.1021/acs.chemrev.5b00744} {\bibfield  {journal} {\bibinfo  {journal}
  {Chem. Rev.}\ }\textbf {\bibinfo {volume} {116}},\ \bibinfo {pages} {7078}
  (\bibinfo {year} {2016})}\BibitemShut {NoStop}%
\bibitem [{\citenamefont {Anwar}\ and\ \citenamefont {Zahn}(2011)}]{Anwar2011}%
  \BibitemOpen
  \bibfield  {author} {\bibinfo {author} {\bibfnamefont {J.}~\bibnamefont
  {Anwar}}\ and\ \bibinfo {author} {\bibfnamefont {D.}~\bibnamefont {Zahn}},\
  }\href@noop {} {\bibfield  {journal} {\bibinfo  {journal} {Angew. Chem. Int.
  Ed.}\ }\textbf {\bibinfo {volume} {50}},\ \bibinfo {pages} {1996} (\bibinfo
  {year} {2011})}\BibitemShut {NoStop}%
\bibitem [{\citenamefont {Jungblut}\ and\ \citenamefont
  {Dellago}(2016)}]{Jungblut2016}%
  \BibitemOpen
  \bibfield  {author} {\bibinfo {author} {\bibfnamefont {S.}~\bibnamefont
  {Jungblut}}\ and\ \bibinfo {author} {\bibfnamefont {C.}~\bibnamefont
  {Dellago}},\ }\href {\doibase 10.1021/acs.chemrev.5b00744} {\bibfield
  {journal} {\bibinfo  {journal} {Eur. Phys. J. E}\ }\textbf {\bibinfo {volume}
  {39}},\ \bibinfo {pages} {77} (\bibinfo {year} {2016})}\BibitemShut {NoStop}%
\bibitem [{\citenamefont {Becker}\ and\ \citenamefont
  {D\"{o}ring}(1935)}]{Becker1935}%
  \BibitemOpen
  \bibfield  {author} {\bibinfo {author} {\bibfnamefont {R.}~\bibnamefont
  {Becker}}\ and\ \bibinfo {author} {\bibfnamefont {W.}~\bibnamefont
  {D\"{o}ring}},\ }\href@noop {} {\bibfield  {journal} {\bibinfo  {journal}
  {Ann. Phys.}\ }\textbf {\bibinfo {volume} {416}},\ \bibinfo {pages} {719}
  (\bibinfo {year} {1935})}\BibitemShut {NoStop}%
\bibitem [{\citenamefont {Binder}(1987)}]{Binder1987}%
  \BibitemOpen
  \bibfield  {author} {\bibinfo {author} {\bibfnamefont {K.}~\bibnamefont
  {Binder}},\ }\href@noop {} {\bibfield  {journal} {\bibinfo  {journal} {Rep.
  Prog. Phys.}\ }\textbf {\bibinfo {volume} {50}},\ \bibinfo {pages} {783}
  (\bibinfo {year} {1987})}\BibitemShut {NoStop}%
\bibitem [{\citenamefont {Moroni}\ \emph {et~al.}(2005)\citenamefont {Moroni},
  \citenamefont {ten Wolde},\ and\ \citenamefont {Bolhuis}}]{Moroni2005}%
  \BibitemOpen
  \bibfield  {author} {\bibinfo {author} {\bibfnamefont {D.}~\bibnamefont
  {Moroni}}, \bibinfo {author} {\bibfnamefont {P.~R.}\ \bibnamefont {ten
  Wolde}}, \ and\ \bibinfo {author} {\bibfnamefont {P.~G.}\ \bibnamefont
  {Bolhuis}},\ }\href@noop {} {\bibfield  {journal} {\bibinfo  {journal} {Phys.
  Rev. Lett.}\ }\textbf {\bibinfo {volume} {94}},\ \bibinfo {pages} {235703}
  (\bibinfo {year} {2005})}\BibitemShut {NoStop}%
\bibitem [{\citenamefont {Trudu}\ \emph {et~al.}(2006)\citenamefont {Trudu},
  \citenamefont {Donadio},\ and\ \citenamefont {Parrinello}}]{Trudu2006}%
  \BibitemOpen
  \bibfield  {author} {\bibinfo {author} {\bibfnamefont {F.}~\bibnamefont
  {Trudu}}, \bibinfo {author} {\bibfnamefont {D.}~\bibnamefont {Donadio}}, \
  and\ \bibinfo {author} {\bibfnamefont {M.}~\bibnamefont {Parrinello}},\
  }\href@noop {} {\bibfield  {journal} {\bibinfo  {journal} {Phys. Rev. Lett.}\
  }\textbf {\bibinfo {volume} {97}},\ \bibinfo {pages} {105701} (\bibinfo
  {year} {2006})}\BibitemShut {NoStop}%
\bibitem [{\citenamefont {Lechner}\ \emph
  {et~al.}(2011{\natexlab{a}})\citenamefont {Lechner}, \citenamefont
  {Dellago},\ and\ \citenamefont {Bolhuis}}]{Lechner2011}%
  \BibitemOpen
  \bibfield  {author} {\bibinfo {author} {\bibfnamefont {W.}~\bibnamefont
  {Lechner}}, \bibinfo {author} {\bibfnamefont {C.}~\bibnamefont {Dellago}}, \
  and\ \bibinfo {author} {\bibfnamefont {P.~G.}\ \bibnamefont {Bolhuis}},\
  }\href@noop {} {\bibfield  {journal} {\bibinfo  {journal} {J. Chem. Phys.}\
  }\textbf {\bibinfo {volume} {135}},\ \bibinfo {pages} {154110} (\bibinfo
  {year} {2011}{\natexlab{a}})}\BibitemShut {NoStop}%
\bibitem [{\citenamefont {Peters}\ and\ \citenamefont
  {Trout}(2006)}]{Peters2006}%
  \BibitemOpen
  \bibfield  {author} {\bibinfo {author} {\bibfnamefont {B.}~\bibnamefont
  {Peters}}\ and\ \bibinfo {author} {\bibfnamefont {B.~L.}\ \bibnamefont
  {Trout}},\ }\href@noop {} {\bibfield  {journal} {\bibinfo  {journal} {J.
  Chem. Phys.}\ }\textbf {\bibinfo {volume} {125}},\ \bibinfo {pages} {054108}
  (\bibinfo {year} {2006})}\BibitemShut {NoStop}%
\bibitem [{\citenamefont {Russo}\ and\ \citenamefont
  {Tanaka}(2016)}]{Tanaka2016}%
  \BibitemOpen
  \bibfield  {author} {\bibinfo {author} {\bibfnamefont {J.}~\bibnamefont
  {Russo}}\ and\ \bibinfo {author} {\bibfnamefont {H.}~\bibnamefont {Tanaka}},\
  }\href {\doibase 10.1063/1.4962166} {\bibfield  {journal} {\bibinfo
  {journal} {J. Chem. Phys.}\ }\textbf {\bibinfo {volume} {145}},\ \bibinfo
  {pages} {211801} (\bibinfo {year} {2016})}\BibitemShut {NoStop}%
\bibitem [{\citenamefont {Wang}\ \emph {et~al.}(2007)\citenamefont {Wang},
  \citenamefont {Gould},\ and\ \citenamefont {Klein}}]{PhysRevE.76.031604}%
  \BibitemOpen
  \bibfield  {author} {\bibinfo {author} {\bibfnamefont {H.}~\bibnamefont
  {Wang}}, \bibinfo {author} {\bibfnamefont {H.}~\bibnamefont {Gould}}, \ and\
  \bibinfo {author} {\bibfnamefont {W.}~\bibnamefont {Klein}},\ }\href
  {\doibase 10.1103/PhysRevE.76.031604} {\bibfield  {journal} {\bibinfo
  {journal} {Phys. Rev. E}\ }\textbf {\bibinfo {volume} {76}},\ \bibinfo
  {pages} {031604} (\bibinfo {year} {2007})}\BibitemShut {NoStop}%
\bibitem [{\citenamefont {Beckham}\ and\ \citenamefont
  {Peters}(2011)}]{Beckham2011}%
  \BibitemOpen
  \bibfield  {author} {\bibinfo {author} {\bibfnamefont {G.~T.}\ \bibnamefont
  {Beckham}}\ and\ \bibinfo {author} {\bibfnamefont {B.}~\bibnamefont
  {Peters}},\ }\href@noop {} {\bibfield  {journal} {\bibinfo  {journal} {J.
  Phys. Chem. Lett.}\ }\textbf {\bibinfo {volume} {2}},\ \bibinfo {pages}
  {1133} (\bibinfo {year} {2011})}\BibitemShut {NoStop}%
\bibitem [{\citenamefont {Jungblut}\ and\ \citenamefont
  {Dellago}(2013)}]{Jungblut2013}%
  \BibitemOpen
  \bibfield  {author} {\bibinfo {author} {\bibfnamefont {S.}~\bibnamefont
  {Jungblut}}\ and\ \bibinfo {author} {\bibfnamefont {C.}~\bibnamefont
  {Dellago}},\ }\href@noop {} {\bibfield  {journal} {\bibinfo  {journal} {Phys.
  Rev. E}\ }\textbf {\bibinfo {volume} {87}},\ \bibinfo {pages} {1} (\bibinfo
  {year} {2013})}\BibitemShut {NoStop}%
\bibitem [{\citenamefont {ten Wolde}\ and\ \citenamefont
  {Frenkel}(1997)}]{tenWolde1997}%
  \BibitemOpen
  \bibfield  {author} {\bibinfo {author} {\bibfnamefont {P.~R.}\ \bibnamefont
  {ten Wolde}}\ and\ \bibinfo {author} {\bibfnamefont {D.}~\bibnamefont
  {Frenkel}},\ }\href@noop {} {\bibfield  {journal} {\bibinfo  {journal}
  {Science}\ }\textbf {\bibinfo {volume} {277}},\ \bibinfo {pages} {1975}
  (\bibinfo {year} {1997})}\BibitemShut {NoStop}%
\bibitem [{\citenamefont {ten Wolde}\ and\ \citenamefont
  {Frenkel}(1999)}]{tenWolde1999}%
  \BibitemOpen
  \bibfield  {author} {\bibinfo {author} {\bibfnamefont {P.~R.}\ \bibnamefont
  {ten Wolde}}\ and\ \bibinfo {author} {\bibfnamefont {D.}~\bibnamefont
  {Frenkel}},\ }\href@noop {} {\bibfield  {journal} {\bibinfo  {journal} {Phys.
  Chem. Chem. Phys.}\ }\textbf {\bibinfo {volume} {1}},\ \bibinfo {pages}
  {2191} (\bibinfo {year} {1999})}\BibitemShut {NoStop}%
\bibitem [{\citenamefont {Schilling}\ \emph {et~al.}(2010)\citenamefont
  {Schilling}, \citenamefont {Sch\"ope}, \citenamefont {Oettel}, \citenamefont
  {Opletal},\ and\ \citenamefont {Snook}}]{PhysRevLett.105.025701}%
  \BibitemOpen
  \bibfield  {author} {\bibinfo {author} {\bibfnamefont {T.}~\bibnamefont
  {Schilling}}, \bibinfo {author} {\bibfnamefont {H.~J.}\ \bibnamefont
  {Sch\"ope}}, \bibinfo {author} {\bibfnamefont {M.}~\bibnamefont {Oettel}},
  \bibinfo {author} {\bibfnamefont {G.}~\bibnamefont {Opletal}}, \ and\
  \bibinfo {author} {\bibfnamefont {I.}~\bibnamefont {Snook}},\ }\href
  {\doibase 10.1103/PhysRevLett.105.025701} {\bibfield  {journal} {\bibinfo
  {journal} {Phys. Rev. Lett.}\ }\textbf {\bibinfo {volume} {105}},\ \bibinfo
  {pages} {025701} (\bibinfo {year} {2010})}\BibitemShut {NoStop}%
\bibitem [{\citenamefont {Kawasaki}\ and\ \citenamefont
  {Tanaka}(2011)}]{Kawasaki2011}%
  \BibitemOpen
  \bibfield  {author} {\bibinfo {author} {\bibfnamefont {T.}~\bibnamefont
  {Kawasaki}}\ and\ \bibinfo {author} {\bibfnamefont {H.}~\bibnamefont
  {Tanaka}},\ }\href@noop {} {\bibfield  {journal} {\bibinfo  {journal} {Proc.
  Natl. Acad. Sci. USA}\ }\textbf {\bibinfo {volume} {108}},\ \bibinfo {pages}
  {6335} (\bibinfo {year} {2011})}\BibitemShut {NoStop}%
\bibitem [{\citenamefont {Russo}\ and\ \citenamefont
  {Tanaka}(2012{\natexlab{a}})}]{Russo2012}%
  \BibitemOpen
  \bibfield  {author} {\bibinfo {author} {\bibfnamefont {J.}~\bibnamefont
  {Russo}}\ and\ \bibinfo {author} {\bibfnamefont {H.}~\bibnamefont {Tanaka}},\
  }\href@noop {} {\bibfield  {journal} {\bibinfo  {journal} {Sci. Rep.}\
  }\textbf {\bibinfo {volume} {2}},\ \bibinfo {pages} {505} (\bibinfo {year}
  {2012}{\natexlab{a}})}\BibitemShut {NoStop}%
\bibitem [{\citenamefont {Russo}\ and\ \citenamefont
  {Tanaka}(2012{\natexlab{b}})}]{Russo2012b}%
  \BibitemOpen
  \bibfield  {author} {\bibinfo {author} {\bibfnamefont {J.}~\bibnamefont
  {Russo}}\ and\ \bibinfo {author} {\bibfnamefont {H.}~\bibnamefont {Tanaka}},\
  }\href@noop {} {\bibfield  {journal} {\bibinfo  {journal} {Soft Matter}\
  }\textbf {\bibinfo {volume} {8}},\ \bibinfo {pages} {4206} (\bibinfo {year}
  {2012}{\natexlab{b}})}\BibitemShut {NoStop}%
\bibitem [{\citenamefont {Leines}\ \emph {et~al.}(2017)\citenamefont {Leines},
  \citenamefont {Drautz},\ and\ \citenamefont {Rogal}}]{DLeines2017}%
  \BibitemOpen
  \bibfield  {author} {\bibinfo {author} {\bibfnamefont {G.~D.}\ \bibnamefont
  {Leines}}, \bibinfo {author} {\bibfnamefont {R.}~\bibnamefont {Drautz}}, \
  and\ \bibinfo {author} {\bibfnamefont {J.}~\bibnamefont {Rogal}},\ }\href
  {\doibase 10.1063/1.4980082} {\bibfield  {journal} {\bibinfo  {journal} {J.
  Chem. Phys.}\ }\textbf {\bibinfo {volume} {146}},\ \bibinfo {pages} {154702}
  (\bibinfo {year} {2017})}\BibitemShut {NoStop}%
\bibitem [{\citenamefont {Lechner}\ \emph
  {et~al.}(2011{\natexlab{b}})\citenamefont {Lechner}, \citenamefont
  {Dellago},\ and\ \citenamefont {Bolhuis}}]{Lechner2011a}%
  \BibitemOpen
  \bibfield  {author} {\bibinfo {author} {\bibfnamefont {W.}~\bibnamefont
  {Lechner}}, \bibinfo {author} {\bibfnamefont {C.}~\bibnamefont {Dellago}}, \
  and\ \bibinfo {author} {\bibfnamefont {P.~G.}\ \bibnamefont {Bolhuis}},\
  }\href@noop {} {\bibfield  {journal} {\bibinfo  {journal} {Phys. Rev. Lett.}\
  }\textbf {\bibinfo {volume} {106}},\ \bibinfo {pages} {1} (\bibinfo {year}
  {2011}{\natexlab{b}})}\BibitemShut {NoStop}%
\bibitem [{\citenamefont {Dellago}\ \emph {et~al.}(2002)\citenamefont
  {Dellago}, \citenamefont {Bolhuis},\ and\ \citenamefont
  {Geissler}}]{Dellago2002}%
  \BibitemOpen
  \bibfield  {author} {\bibinfo {author} {\bibfnamefont {C.}~\bibnamefont
  {Dellago}}, \bibinfo {author} {\bibfnamefont {P.}~\bibnamefont {Bolhuis}}, \
  and\ \bibinfo {author} {\bibfnamefont {P.~L.}\ \bibnamefont {Geissler}},\
  }\href@noop {} {\bibfield  {journal} {\bibinfo  {journal} {Adv. Chem. Phys.}\
  }\textbf {\bibinfo {volume} {123}},\ \bibinfo {pages} {1} (\bibinfo {year}
  {2002})}\BibitemShut {NoStop}%
\bibitem [{\citenamefont {van Erp}\ and\ \citenamefont
  {Bolhuis}(2005)}]{VanErp2005}%
  \BibitemOpen
  \bibfield  {author} {\bibinfo {author} {\bibfnamefont {T.~S.}\ \bibnamefont
  {van Erp}}\ and\ \bibinfo {author} {\bibfnamefont {P.~G.}\ \bibnamefont
  {Bolhuis}},\ }\href@noop {} {\bibfield  {journal} {\bibinfo  {journal} {J.
  Comp. Phys.}\ }\textbf {\bibinfo {volume} {205}},\ \bibinfo {pages} {157}
  (\bibinfo {year} {2005})}\BibitemShut {NoStop}%
\bibitem [{\citenamefont {Rogal}\ \emph {et~al.}(2010)\citenamefont {Rogal},
  \citenamefont {Lechner}, \citenamefont {Juraszek}, \citenamefont {Ensing},\
  and\ \citenamefont {Bolhuis}}]{Rogal2010}%
  \BibitemOpen
  \bibfield  {author} {\bibinfo {author} {\bibfnamefont {J.}~\bibnamefont
  {Rogal}}, \bibinfo {author} {\bibfnamefont {W.}~\bibnamefont {Lechner}},
  \bibinfo {author} {\bibfnamefont {J.}~\bibnamefont {Juraszek}}, \bibinfo
  {author} {\bibfnamefont {B.}~\bibnamefont {Ensing}}, \ and\ \bibinfo {author}
  {\bibfnamefont {P.~G.}\ \bibnamefont {Bolhuis}},\ }\href@noop {} {\bibfield
  {journal} {\bibinfo  {journal} {J. Chem. Phys.}\ }\textbf {\bibinfo {volume}
  {133}},\ \bibinfo {pages} {174109} (\bibinfo {year} {2010})}\BibitemShut
  {NoStop}%
\bibitem [{\citenamefont {Lechner}\ \emph {et~al.}(2010)\citenamefont
  {Lechner}, \citenamefont {Rogal}, \citenamefont {Juraszek}, \citenamefont
  {Ensing},\ and\ \citenamefont {Bolhuis}}]{Lechner2010}%
  \BibitemOpen
  \bibfield  {author} {\bibinfo {author} {\bibfnamefont {W.}~\bibnamefont
  {Lechner}}, \bibinfo {author} {\bibfnamefont {J.}~\bibnamefont {Rogal}},
  \bibinfo {author} {\bibfnamefont {J.}~\bibnamefont {Juraszek}}, \bibinfo
  {author} {\bibfnamefont {B.}~\bibnamefont {Ensing}}, \ and\ \bibinfo {author}
  {\bibfnamefont {P.~G.}\ \bibnamefont {Bolhuis}},\ }\href {\doibase
  10.1063/1.3491818} {\bibfield  {journal} {\bibinfo  {journal} {J. Chem.
  Phys.}\ }\textbf {\bibinfo {volume} {133}},\ \bibinfo {pages} {174110}
  (\bibinfo {year} {2010})}\BibitemShut {NoStop}%
\bibitem [{\citenamefont {Bokeloh}\ \emph {et~al.}(2011)\citenamefont
  {Bokeloh}, \citenamefont {Rozas}, \citenamefont {Horbach},\ and\
  \citenamefont {Wilde}}]{Bokeloh2011}%
  \BibitemOpen
  \bibfield  {author} {\bibinfo {author} {\bibfnamefont {J.}~\bibnamefont
  {Bokeloh}}, \bibinfo {author} {\bibfnamefont {R.~E.}\ \bibnamefont {Rozas}},
  \bibinfo {author} {\bibfnamefont {J.}~\bibnamefont {Horbach}}, \ and\
  \bibinfo {author} {\bibfnamefont {G.}~\bibnamefont {Wilde}},\ }\href@noop {}
  {\bibfield  {journal} {\bibinfo  {journal} {Phys. Rev. Lett.}\ }\textbf
  {\bibinfo {volume} {107}},\ \bibinfo {pages} {1} (\bibinfo {year}
  {2011})}\BibitemShut {NoStop}%
\bibitem [{\citenamefont {Bolhuis}\ and\ \citenamefont
  {Lechner}(2011)}]{Bolhuis2011}%
  \BibitemOpen
  \bibfield  {author} {\bibinfo {author} {\bibfnamefont {P.~G.}\ \bibnamefont
  {Bolhuis}}\ and\ \bibinfo {author} {\bibfnamefont {W.}~\bibnamefont
  {Lechner}},\ }\href@noop {} {\bibfield  {journal} {\bibinfo  {journal} {J.
  Stat. Phys.}\ }\textbf {\bibinfo {volume} {145}},\ \bibinfo {pages} {841}
  (\bibinfo {year} {2011})}\BibitemShut {NoStop}%
\bibitem [{\citenamefont {Dellago}\ \emph {et~al.}(1998)\citenamefont
  {Dellago}, \citenamefont {Bolhuis}, \citenamefont {Csajka},\ and\
  \citenamefont {Chandler}}]{Dellago1998}%
  \BibitemOpen
  \bibfield  {author} {\bibinfo {author} {\bibfnamefont {C.}~\bibnamefont
  {Dellago}}, \bibinfo {author} {\bibfnamefont {P.~G.}\ \bibnamefont
  {Bolhuis}}, \bibinfo {author} {\bibfnamefont {F.~S.}\ \bibnamefont {Csajka}},
  \ and\ \bibinfo {author} {\bibfnamefont {D.}~\bibnamefont {Chandler}},\
  }\href@noop {} {\bibfield  {journal} {\bibinfo  {journal} {J. Chem. Phys.}\
  }\textbf {\bibinfo {volume} {108}},\ \bibinfo {pages} {1964} (\bibinfo {year}
  {1998})}\BibitemShut {NoStop}%
\bibitem [{\citenamefont {van Erp}(2007)}]{VanErp2007}%
  \BibitemOpen
  \bibfield  {author} {\bibinfo {author} {\bibfnamefont {T.~S.}\ \bibnamefont
  {van Erp}},\ }\href@noop {} {\bibfield  {journal} {\bibinfo  {journal} {Phys.
  Rev. Lett.}\ }\textbf {\bibinfo {volume} {98}},\ \bibinfo {pages} {268301}
  (\bibinfo {year} {2007})}\BibitemShut {NoStop}%
\bibitem [{\citenamefont {Bolhuis}(2008)}]{Bolhuis2008}%
  \BibitemOpen
  \bibfield  {author} {\bibinfo {author} {\bibfnamefont {P.~G.}\ \bibnamefont
  {Bolhuis}},\ }\href@noop {} {\bibfield  {journal} {\bibinfo  {journal} {J.
  Chem. Phys.}\ }\textbf {\bibinfo {volume} {129}},\ \bibinfo {pages} {114108}
  (\bibinfo {year} {2008})}\BibitemShut {NoStop}%
\bibitem [{\citenamefont {Ferrenberg}\ and\ \citenamefont
  {Swendsen}(1989)}]{Ferrenberg1989}%
  \BibitemOpen
  \bibfield  {author} {\bibinfo {author} {\bibfnamefont {A.~M.}\ \bibnamefont
  {Ferrenberg}}\ and\ \bibinfo {author} {\bibfnamefont {R.~H.}\ \bibnamefont
  {Swendsen}},\ }\href {\doibase 10.1103/PhysRevLett.63.1195} {\bibfield
  {journal} {\bibinfo  {journal} {Phys. Rev. Lett.}\ }\textbf {\bibinfo
  {volume} {63}},\ \bibinfo {pages} {1195} (\bibinfo {year}
  {1989})}\BibitemShut {NoStop}%
\bibitem [{\citenamefont {Husmeier}(2005)}]{Husmeier2005}%
  \BibitemOpen
  \bibfield  {author} {\bibinfo {author} {\bibfnamefont {D.}~\bibnamefont
  {Husmeier}},\ }in\ \href@noop {} {\emph {\bibinfo {booktitle} {Probabilistic
  Modeling in Bioinformatics and Medical Informatics}}},\ \bibinfo {editor}
  {edited by\ \bibinfo {editor} {\bibfnamefont {D.}~\bibnamefont {Husmeier}},
  \bibinfo {editor} {\bibfnamefont {R.}~\bibnamefont {Dybowski}}, \ and\
  \bibinfo {editor} {\bibfnamefont {S.}~\bibnamefont {Roberts}}}\ (\bibinfo
  {publisher} {Springer},\ \bibinfo {address} {London},\ \bibinfo {year}
  {2005})\ p.~\bibinfo {pages} {17}\BibitemShut {NoStop}%
\bibitem [{\citenamefont {Peters}\ \emph {et~al.}(2007)\citenamefont {Peters},
  \citenamefont {Beckham},\ and\ \citenamefont {Trout}}]{Peters2007}%
  \BibitemOpen
  \bibfield  {author} {\bibinfo {author} {\bibfnamefont {B.}~\bibnamefont
  {Peters}}, \bibinfo {author} {\bibfnamefont {G.~T.}\ \bibnamefont {Beckham}},
  \ and\ \bibinfo {author} {\bibfnamefont {B.~L.}\ \bibnamefont {Trout}},\
  }\href@noop {} {\bibfield  {journal} {\bibinfo  {journal} {J. Chem. Phys.}\
  }\textbf {\bibinfo {volume} {127}},\ \bibinfo {pages} {034109} (\bibinfo
  {year} {2007})}\BibitemShut {NoStop}%
\bibitem [{\citenamefont {Schwarz}(1978)}]{Schwarz1978}%
  \BibitemOpen
  \bibfield  {author} {\bibinfo {author} {\bibfnamefont {G.}~\bibnamefont
  {Schwarz}},\ }\href@noop {} {\bibfield  {journal} {\bibinfo  {journal} {Ann.
  Stat.}\ }\textbf {\bibinfo {volume} {6}},\ \bibinfo {pages} {461} (\bibinfo
  {year} {1978})}\BibitemShut {NoStop}%
\bibitem [{\citenamefont {Foiles}\ \emph {et~al.}(1986)\citenamefont {Foiles},
  \citenamefont {Baskes},\ and\ \citenamefont {Daw}}]{Foiles86}%
  \BibitemOpen
  \bibfield  {author} {\bibinfo {author} {\bibfnamefont {S.~M.}\ \bibnamefont
  {Foiles}}, \bibinfo {author} {\bibfnamefont {M.~I.}\ \bibnamefont {Baskes}},
  \ and\ \bibinfo {author} {\bibfnamefont {M.~S.}\ \bibnamefont {Daw}},\ }\href
  {\doibase 10.1103/PhysRevB.33.7983} {\bibfield  {journal} {\bibinfo
  {journal} {Phys. Rev. B}\ }\textbf {\bibinfo {volume} {33}},\ \bibinfo
  {pages} {7983} (\bibinfo {year} {1986})}\BibitemShut {NoStop}%
\bibitem [{\citenamefont {Plimpton}(1995)}]{Plimpton1995}%
  \BibitemOpen
  \bibfield  {author} {\bibinfo {author} {\bibfnamefont {S.}~\bibnamefont
  {Plimpton}},\ }\href@noop {} {\bibfield  {journal} {\bibinfo  {journal} {J.
  Comp. Phys.}\ }\textbf {\bibinfo {volume} {117}},\ \bibinfo {pages} {1}
  (\bibinfo {year} {1995})}\BibitemShut {NoStop}%
\bibitem [{\citenamefont {Steinhardt}\ \emph {et~al.}(1983)\citenamefont
  {Steinhardt}, \citenamefont {Nelson},\ and\ \citenamefont
  {Ronchetti}}]{Steinhardt1983}%
  \BibitemOpen
  \bibfield  {author} {\bibinfo {author} {\bibfnamefont {P.~J.}\ \bibnamefont
  {Steinhardt}}, \bibinfo {author} {\bibfnamefont {D.~R.}\ \bibnamefont
  {Nelson}}, \ and\ \bibinfo {author} {\bibfnamefont {M.}~\bibnamefont
  {Ronchetti}},\ }\href@noop {} {\bibfield  {journal} {\bibinfo  {journal}
  {Phys. Rev. B}\ }\textbf {\bibinfo {volume} {28}},\ \bibinfo {pages} {784}
  (\bibinfo {year} {1983})}\BibitemShut {NoStop}%
\bibitem [{\citenamefont {Auer}\ and\ \citenamefont
  {Frenkel}(2005)}]{Auer2005}%
  \BibitemOpen
  \bibfield  {author} {\bibinfo {author} {\bibfnamefont {S.}~\bibnamefont
  {Auer}}\ and\ \bibinfo {author} {\bibfnamefont {D.}~\bibnamefont {Frenkel}},\
  }\href@noop {} {\bibfield  {journal} {\bibinfo  {journal} {Adv. Polym. Sci.}\
  }\textbf {\bibinfo {volume} {173}},\ \bibinfo {pages} {149} (\bibinfo {year}
  {2005})}\BibitemShut {NoStop}%
\bibitem [{\citenamefont {Lechner}\ and\ \citenamefont
  {Dellago}(2008)}]{Lechner2008}%
  \BibitemOpen
  \bibfield  {author} {\bibinfo {author} {\bibfnamefont {W.}~\bibnamefont
  {Lechner}}\ and\ \bibinfo {author} {\bibfnamefont {C.}~\bibnamefont
  {Dellago}},\ }\href@noop {} {\bibfield  {journal} {\bibinfo  {journal} {J.
  Chem. Phys.}\ }\textbf {\bibinfo {volume} {129}},\ \bibinfo {pages} {114707}
  (\bibinfo {year} {2008})}\BibitemShut {NoStop}%
\bibitem [{\citenamefont {Herlach}\ \emph {et~al.}(2016)\citenamefont
  {Herlach}, \citenamefont {Palberg}, \citenamefont {Klassen}, \citenamefont
  {Klein},\ and\ \citenamefont {Kobold}}]{Herlach2016}%
  \BibitemOpen
  \bibfield  {author} {\bibinfo {author} {\bibfnamefont {D.~M.}\ \bibnamefont
  {Herlach}}, \bibinfo {author} {\bibfnamefont {T.}~\bibnamefont {Palberg}},
  \bibinfo {author} {\bibfnamefont {I.}~\bibnamefont {Klassen}}, \bibinfo
  {author} {\bibfnamefont {S.}~\bibnamefont {Klein}}, \ and\ \bibinfo {author}
  {\bibfnamefont {R.}~\bibnamefont {Kobold}},\ }\href@noop {} {\bibfield
  {journal} {\bibinfo  {journal} {J. Chem. Phys.}\ }\textbf {\bibinfo {volume}
  {145}},\ \bibinfo {pages} {211703} (\bibinfo {year} {2016})}\BibitemShut
  {NoStop}%
\bibitem [{\citenamefont {Schablitzki}\ \emph {et~al.}(2013)\citenamefont
  {Schablitzki}, \citenamefont {Rogal},\ and\ \citenamefont
  {Drautz}}]{Schablitzki2013}%
  \BibitemOpen
  \bibfield  {author} {\bibinfo {author} {\bibfnamefont {T.}~\bibnamefont
  {Schablitzki}}, \bibinfo {author} {\bibfnamefont {J.}~\bibnamefont {Rogal}},
  \ and\ \bibinfo {author} {\bibfnamefont {R.}~\bibnamefont {Drautz}},\
  }\href@noop {} {\bibfield  {journal} {\bibinfo  {journal} {Modelling Simul.
  Mater. Sci. Eng.}\ }\textbf {\bibinfo {volume} {21}},\ \bibinfo {pages}
  {075008} (\bibinfo {year} {2013})}\BibitemShut {NoStop}%
\end{thebibliography}%

%

\end{document}